# Application of lower temperature for crystallization of PbTiO3 nanopowders by the Sol-gel method

Maryam Lanki[1], Abolghasem Nourmohammadi[2*], Seyed Mohammad Hassan Feiz[1], Ehsan Rahmati Adarmanabadi[1]

[1] Department of Physics, Faculty of Science, University of Isfahan, 81746-73441, Isfahan, Iran
[2] Department of Nanotechnology, Faculty of Advanced Science and Technologies, University of Isfahan, 81746-73441, Isfahan, Iran
*Corresponding Author, E-mail: a.nourmohammadi@sci.ui.ac.ir
Phone: +98-31-37934404
Fax: +98-31-37932342

**Abstract—**

The aim of this article is to decrease the crystallization temperature of the tetragonal perovskite $PbTiO_3$ nanopowders below the $PbTiO_3$ Curie temperature. At the beginning of our study, the calcination temperature was selected based on the thermal analysis of the $PbTiO_3$ gels, came out at 680°C. However, reconsidering the thermal properties of the $PbTiO_3$ gel revealed the possibility of the formation of single tetragonal perovskite phase at lower temperatures. Both X-Ray diffraction and transmission electron microscopy investigations showed that well crystalline pure phase $PbTiO_3$ nanopowders have been synthesized after calcination of the $PbTiO_3$ gel at 460°C, which is quite to the Curie temperature of the $PbTiO_3$ nanoparticles. To improve crystallization of the samples prepared at 460°C, the effect of the heat treatment parameters and Pb content on gel crystallization and the Pb-partitioning phenomenon (section 3.2) were investigated in detail.

**Keywords-** Nanopowders, $PbTiO_3$, Sol-gel Preparation, Crystallization.

**1.    Introduction**

Lead titanate or $PbTiO_3$, with the tetragonal perovskite structure, is a well-known piezoelectric and ferroelectric material with good piezoelectric and ferroelectric properties. This material exhibits a wide spectrum of functional properties such as switchable polarization, piezoelectricity, high non-linear optical activity, pyroelectricity and high dielectric behavior [1, 2]. However, similar to other PbO-based ferroelectrics, PbTiO3 has several technical problems which limit its fabrication and application. Since PbO is a toxic volatile material, its evaporation at high temperatures produces porosity in the synthesized material and makes poison productions or releases toxic Pb compounds, which poses a hazard for the environment.

A specific drawback of PbTiO3 is its very large lattice distortion during cubic- tetragonal phase transition around 500°C. It is difficult to fabricate high quality PbTiO3 ceramics above 500°C because of large lattice distortion due to cubic- tetragonal phase transformation about 500°C. Large volume change upon cooling around this temperature leads to internal stresses in the samples. For this reason, bulk sintered PbTiO3 ceramics are fragile and may disintegrate into powder at room temperature due to mechanical breakdown[3]. For several decades, researchers have attempted fabricating dense modified PbTiO3 ceramics by adding suitable additives. However, it is known that most of these additives change the ferroelectric properties of PbTiO3[4, 5].

Some authors have already reported various methods of preparation of PbTiO3 nanopowders such as conventional mixed-oxide, Pechini-type processes, mechanochemical synthesis, hydrothermal process, sputtering, spray drying and sol-gel processing [6, 7, 8, 9, 10, 11]. Among them, the sol-gel methods have the advantages of being cheap and simple and precise control of the composition.

To the best of our knowledge, the entire crystallization of the tetragonal perovskite $PbTiO_3$ nanopowder by the sol-gel method requires heat treatment of the PbTiO3 gel between 650°C and 700°C[10, 11, 12, 13]. Such crystallization temperature range is a disadvantage as it results in an increase of the crystallite size as well as an increase of Pb evaporation. In addition, PbTiO3 nanostructures which are sintered at these temperatures (650-700°C) face up to a large lattice distortion due to cubic- tetragonal phase transformation about 500°C. Therefore, many attempts have been already done to decrease the crystallization temperature of PbTiO3 nanopowders produced by sol-gel processing. This particular topic is considered by us too. We have investigated the Pb-partitioning phenomenon in more detail in PbTiO3 nanopowders in order to reach the crystallization below 500°C. Our approach has the advantage of not suffering from the disadvantages attributed to the conventional processing of Pb-based ferroelectric materials via the sol-gel method.

## 2. Experimental

### 2.1 Preparing PbTiO$_3$ precursor sols

Highly stabilized $PbTiO_3$ precursor sols were prepared through chemical modification of the organometallic precursors of titanium by acetic acid. Details of this method was already reported elsewhere [10]. Typically, in order to prepare the $Pb_{1.05}TiO_3$ precursor sol, 27.95gr lead acetate trihydrate (99.5% purity, Merck) was dissolved in 24.04ml pure glacial acetic acid. Then, the prepared solution was dehydrated at 110ºC for 15min and cooled down to room temperature. After cooling down, 23.25ml titanium butoxide (98% purity, Merck) was added gradually, drop by drop, to the solution. For 15min, this solution was stirred at room temperature, and then, a mixture of 12.01ml de-ionized water and 33.45ml ethanol (99.5% purity, Merck) was added to cause hydrolysis and prevent fast gelation of the prepared sol. Then, 4.13ml Ethylene glycol (99% purity, Merck) and 6.92ml acetyl acetone (99% purity, Merck) were added to enhance viscosity and stability of the final precursor solution.

## 2.2    Preparation and characterization of PbTiO$_3$ nanopowders

The prepared sols were heated, along with continuous stirring, to obtain the PbTiO$_3$ gels, and the prepared gel was subsequently dried at 100ºC for 10 hours. Then, the gel calcination temperature for the growth of the perovskite tetragonal structure in the PbTiO$_3$ powder was selected based on simultaneous thermal analysis (STA) of the prepared gel. Thermal analysis was carried out on a Perkin Elmer model Pyris diamond S(II) thermal analyzer with the rating of 10°C/min.

The prepared PbTiO$_3$ nanopowders were investigated using a transmission electron microscope (TEM) system model Philips EM208S. Then, the nanopowders were analyzed by X-Ray diffraction (XRD) employing a Bruker-d8 advance model using Cu Kα radiation in order to identify the crystal structure of the calcinated PbTiO$_3$ nanopowders, calculating the crystallized percentage of the powder semi-quantitatively, and  evaluating  the best heat treatment condition for preparing well crystallized homogenous PbTiO$_3$ nanopowders. XRD data were recorded using fix angular and time steps and reported as measured (i.e., no data normalization were performed).

## 3.    Results and discussion

### 3.1 Reconsidering thermal properties of the PbTiO$_3$ gel

Figure 1 shows the results of the STA analysis of the prepared Pb$_{1.05}$TiO$_3$ gel. As it is seen, at the beginning of the DTA graph, a broad exothermic region is observed below 280°C (Point O). This exothermic region is accompanied by a steep decrease of the weight in the thermal gravimetry (TG) plot (the AB region in Figure 1). Therefore, it is due to the starting decomposition of the organic materials and evaporation of the solvent. Afterwards, another exothermic region is observed in the DTA graph below 320°C (Point P), and an endothermic region is located around 330°C (Point Q), which the latter is accompanied by a steep decrease of the weight in the TG plot (the BC region, Figure 1). Then a very broad weak exothermic region is observed below 400°C (Point R). Weight loss is observed within the 330- 400°C range, but, rate of the weight reduction rate decreases in this region (i.e. the BC region), so that, the weight loss in this region may be due to the endothermic region about 330°C (Point Q). Therefore, the 330- 400°C range could be considered as the crystallization range for the perovskite or an intermediate $PbTiO_3$ phase. Furthermore, there is a rather sharp exothermic peak around 460°C (Point S), which is related to the removal of the rest of the left organic materials as well as the structural water. Later, a wide exothermic peak was seen at 680°C in the DTA graph (Point S). No weight loss was observed in the TG graph around this temperature (the DE region, Figure 1). Therefore, it was concluded that the mentioned peak (680°C) is due to crystallization of the perovskite $PbTiO_3$. Based on the latter results, heat treatment at 680°C was successfully tested for the preparation of $PbTiO_3$ nanopowders in our previous work [10]. But, here, in order to identify which phase could crystallize in the 330- 400°C temperature range and decrease the crystallization temperature of the tetragonal perovskite $PbTiO_3$ nanopowders below 500°C, the following investigations were carried out at both 460°C and 680°C. Pb-partitioning phenomenon was investigated in PbTiO3 nanopowders produced by sol-gel processing. Moreover, the effects of the heat treatment parameters on the perovskite phase crystallization was studied.

## 3.2 Pb-partitioning phenomenon in the PbTiO3 nanopowders

In the first part of this study, Pb-partitioning phenomenon was investigated in our PbTiO3 nanopowders in order to improve crystallization below 500°C. Different extra Pb contents were added to the $PbTiO_3$ sols in order to improve crystallization. Thus, it was expected that Pb-partitioning phenomenon would be observed in some $PbTiO_3$ nanopowder samples, after heat treatment at 680°C. The Pb-partitioning phenomenon (formation of Pb element during heat treatment) happens in varying amount of containing excess Pb in the PbO-based samples during decomposition of the orgnic materials. Presence of the carbonaceous species provides a reducing atmosphere locally and may reduce PbO into Pb element. This phenomenon makes crystallization of the perovskite phase difficult (time consuming) and leaves toxic Pb compounds in the samples.

After heat treatment at 680°C for 6 hours, XRD patterns of the produced $PbTiO_3$ powders having extra Pb contents of 0%, 5% and 10% (Samples A-C, in table 1) are shown in Figure 2 in the measured scale. It is observed that the tetragonal phase of $PbTiO_3$ compound is formed from all $PbTiO_3$ gels after the aforementioned heat treatment condition. However, crystalline percentage is different in the prepared samples. In addition, there are some peaks associated with the undesirable PbO phase present in the XRD pattern of the annealed $PbTiO_3$ gels with 5% and 10% (sample A and B, Table I) extra lead content.

The undesirable PbO phase has been removed in Figure 1-a (0% extra lead), but the crystallinity of the sample, percentage of the tetragonal perovskite $PbTiO_3$ phase, is the least among the analyzed samples, for the sample containing 0% extra lead (sample C, Table I). XRD pattern of the annealed $PbTiO_3$ gel with 5% extra lead is shown in Figure 2-b. This figure indicates clearly that some amount of the undesired PbO phase exists in the XRD pattern of the sample containing 5% extra lead (sample B, Table I), but the crystalline intensity is more than the sample containing 0% extra lead (sample C, Table I). Moreover, less amount of the undesired PbO phase exists in the XRD pattern, but the crystalline intensity is also less than the sample containing 10% extra lead, see Figure 2-c (sample A, Table I).

It is shown here that Pb content has a key contribution in the crystallization of the PbTiO3 tetragonal perovskite phase. Samples having extra Pb content of 5% were selected for the rest of the current investigation.

### 3.3 Effect of the calcination time

The next parameter which is here studied is, for obtaining crystallization of the tetragonal perovskite $PbTiO_3$ nanopowders below 500°C, was the period of the final calcination step. Figures 3a and 3b shows the effect of the calcination time on the phase structure of the lead titanate powder containing 5%wt extra lead. The XRD powder samples are prepared by applying different time of the heat treating at 680°C for 2 , 4 and 6 hours , respectively (Samples E, D and B in table 1, respectively).

The crystalline percentage of the prepared lead titanate powders annealed at 680°C was also calculated semi-quantitatively using the Bruker-D8 advance model software as reported in table 1.

In both Figure 3b and Figure 3c, the intensity of the peak associated with the (101) plane has increased compared to Figure 3a. However, other diffraction peaks have larger intensities in Figure 3a. It is interesting to note that increasing heat treating time at 680°C for 2 to 6 hours has conversely affected the crystalline percentage of the prepared lead titanate powders (as observed in table 1). Similar effects were also observed in other samples. The reason may be PbO evaporation during the heat treatment stage, as confirmed by the large variation of the XRD peaks due to PbO (Figure 3 a-c). Since PbO content has a key contribution in the crystallization of the PbTiO$_3$ tetragonal perovskite phase (section 3.2), PbO evaporation results in lower PbTiO$_3$ crystallization. Accordingly it is recognized significance of utilizing lower calcination time in order to improve crystallization of the tetragonal perovskite PbTiO$_3$ nanopowders below 500°C. Our results makes makes the fabrication of PbTiO$_3$ nanopowders cost effective, because table 1 shows that more energy consumption for the application of longer heat treatment procedures is not practically reasonable.

**3.4 Crystallization of PbTiO3 nanopowders at lower temperatures**

In order to obtain the crystallization of PbTiO3 nanopowders at lower temperatures, the effect of the firing temperature was also investigated. Figures 4-a and 4-b show the XRD patterns of the PbTiO$_3$ gels, with 5% extra Pb content, which are heat treated at 460°C and 680°C, respectively. Both samples are heated with a similar heating rate of 20°C/min. Crystallization of tetragonal perovskite phase of PbTiO$_3$ was demonstrated by XRD pattern of the powders, see figure 4-a (sample F, Table I) and figure 4-b (sample E, Table I). Our XRD results clearly show that the first temperature for starting the crystallization of lead titanate occurs below 460°C.

It should be noted that we could not get rid of the peaks due to PbO in the samples heat treated at 680°C for 2 hours which result in fabrication of poison PbTiO$_3$ productions, see figures 2-a and 4-b. However, the intensity of the PbO peaks decreased drastically after heat treatment at 460°C. Our results confirm the beneficial effect of the lower temperature crystallization of PbTiO3 nanopowders for the environment.

It is found here that it is possible to minimize the undesirable Pb-partitioning phenomenon in PbTiO$_3$ nanopowders and simultaneously improve crystallization of the PbTiO3 tetragonal perovskite phase at lower temperatures.

### 3.5 Effect of the heating rate

In the fourth set of our experiments, the effects of the heating rate on the crystallization of the tetragonal perovskite PbTiO3 and the final phase composition of the prepared nanopowders are studied.

Figure 5-a and 5-b show the XRD patterns of the PbTiO$_3$ gels with 5% extra Pb content, which are heated at the rate of 10°C/min (sample G, table 1) and 20°C/min (sample F, table 1), respectively. Both samples are heat treated at 460°C, and, as it is seen in table 1, the sample which is heat treated with the 20 ºC/min heating rate exhibits better crystallization of PbTiO$_3$. On the other hand, the sample heat treated with the 10 ºC/min heating rate exhibit formation of more undesirable PbO phase and less crystallization of PbTiO3 perovskite phase. This indicates the existence of more organic components due to less decomposition of the organics in the intermediate temperatures, which results in Pb-partitioning phenomenon (Figure 5-a, sample G in table 1). Hence, a 20 ºC/min heating rate was selected in the rest of our study for better crystallization and lower Pb-partitioning.

Our results may indicate complete removal of the organics about 460°C. However, for entire crystallization of PbTiO3 nanopowders at this temperature it is necessary to optimize the heat treatment parameters.

Finally, the powders prepared with the optimum conditions, heat treated at a lower temperature, 460°C, for 2 hours with a 20 ºC/min heating rate, was selected for our TEM analyses.

It should be noted that, our XRD results showed that the $Pb_{1.00}TiO_3$ gel (gel containing no extra Pb content) shows crystallization of single phase tetragonal $PbTiO_3$ after heat treatment at 680°C (Figure 2a). This temperature is in the same order of our previous studies. However, by adding extra Pb into the precursor sol and studying the Pb-partitioning phenomenon in this material, XRD investigations confirmed that the researchers here have obtained single phase crystallization of lead titanate nanopowders by sol-gel method about 460°C which is a much lower than the temperature range 650-700°C as reported by others[11, 12, 13].

### 3.2 TEM Analysis

TEM microstructural and electron diffraction analyses were performed in order to investigate size, morphology and phase composition of the prepared $PbTiO_3$ nanoparticles. To confirm our XRD results, the $PbTiO_3$ powder containing 5%wt extra lead and heated for 2 hours at 460°C, using a 20°C/min heating rate, was selected. As shown in Figure 6, a uniform morphology is achieved in $PbTiO_3$ nanoparticles heat treated at 460°C, which indicates that our method has been a suitable synthesis one.

Figure 7 shows the diffraction pattern of the $PbTiO_3$ nanoparticles of Figure 6. All the diffraction points correspond with the tetragonal perovskite phase according to the JCPDS card No. 6-452. As shown in this figure, single phase tetragonal perovskite crystallization of lead titanate is formed. It is demonstrated that low temperature single phase crystallization of $PbTiO_3$ has occurred at 460°C whereas crystallization of $PbTiO_3$ has already been reported above 650°C [11, 12, 13].

Figure 7 shows the splitting of the diffraction points in the electron diffraction pattern of the prepared lead titanate nanoparticles. This result clearly indicates tetragonal related distortion of the perovskite structure that had been confirmed with peak splitting in XRD results; see Figs. 4 and 5. It is observed that well crystalline PbTiO3 nanoparticles are prepared at 460°C (Figure 7), while tetragonal perovskite PbTiO3 are already synthesized at 680°C [10].

**3.3 Advantages of our method**

Our XRD results confirm the effect of the lower temperature crystallization of $PbTiO_3$ nanopowders to avoid formation of poison productions or releases toxic Pb compounds into the environment (section 3.4).

As shown by our TEM micrographs, the prepared $PbTiO_3$ nanoparticles mostly have dimension of about 25nm. Thus, according to the empirical expression presented by Ishikawa and coworkers[14], they have a Curie temperature of above 450°C. By using our method, pure tetragonal $PbTiO_3$ nanostructures can be sintered below 460°C. As a result, the $PbTiO_3$ nanostructures which are prepared by this way can avoid the large lattice distortion due to cubic-tetragonal phase transformation.

Many researchers have attempted lower temperature crystallization of $PbTiO_3$ nanopowders. Ishikawa and coworkers were prepared PbTiO3 nanoparticles via the sol-gel process using metal alkoxide precursors[11]. They claimed that crystallization of the amorphous PbTiO3 gel may occur below 460°C. However, they did not present any experimental evidence. Our research works indicate that crystallization of the amorphous PbTiO3 gel into the tetragonal perovskite phase definitely occurs below 460°C by selecting a convenient Pb percentage and an optimized heat treatment process.

Kakihana and coworkers reported crystallization of $PbTiO_3$ nanopowders at 400°C[15]. However, their method is quite different from us. It should be noted that their synthesis method is mostly applicable for powder preparation. But, our alkoxide-based method is more general than their method, because our method is applicable for the synthesis of various nanostructures (thin films, nanodots, nanorods, etc).

## 4. Conclusion

Crystallization temperature is a vital problem in nanotechnology. The researchers here have obtained single phase crystallization of lead titanate nanopowders by the sol-gel method at 460°C which is a much lower temperature than the 650-700°C range, as reported by others. Crystallization at 460°C decreases Pb evaporation efficiently during the heat treatment, which can minimize release of toxic Pb compounds in the environment and can protect the stoichiometric ratio of Pb/Ti.

Our XRD investigations on the calcinated nanopowders showed that the complete crystallization of the amorphous $PbTiO_3$ into the $PbTiO_3$ perovskite structure would be possible only after optimizing the heat treatment condition and investigating the effect of Pb content, as performed here.

Our TEM and electron diffraction investigations showed that well crystalline $PbTiO_3$ nanoparticles with a rather uniform morphology can be produced at 460°C whereas formation of such nanoparticles have already been reported above 650°C.

It was also shown that in the future, by using our method, the $PbTiO_3$ nanostructures can avoid the large lattice distortion due to cubic- tetragonal phase transformation.

Our results demonstrated the essential importance of the lower temperature crystallization of $PbTiO_3$ nanopowders. To the best of our knowledge, our article is the first one which presents experimental evidence on crystallization of the amorphous $PbTiO_3$ gel into the perovskite phase below 460°C, for the PbTiO$_3$ nanoparticles produced by the alkoxide-based sol-gel method. This issue has been attempted by other researchers too. But, no evidence has been already presented. Crystallization of PbTiO$_3$ nanopowders at 400°C is already reported for the polymerized complex synthesis method, which is mostly applicable for powder preparation. But, alkoxide-based method is more general, because this method can be applied for the fabrication of different nanostructures (nanowires, nanotubes, etc).

**Acknowledgment**

The authors would like to thank the Office of Graduate Studies of the University of Isfahan for their support.

14**Figure captions**

**Figure 1.** (Color online) The STA (DTA/TG) analysis of the dried $PbTiO_3$ gel.

**Figure 2.** Pb-partitioning results: XRD patterns of the $PbTiO_3$ gels with the extra Pb contents of a) 0%, b)5% , and c)10% after annealing at 680°C for 6hours.

**Figure 3**. Effect of the calcination time: XRD patterns of the $PbTiO_3$ nanopowders with the extra Pb content of 5% prepared at 680°C for a) 2 hours, b) 4 hours, c)6 hours.

**Figure 4**. Effect of firing temperature on crystallization of $Pb_{1.05}TiO_3$ powder at a) 460°C b) 680°C.

**Figure 5**. Effect of the heating rate: XRD patterns of the $PbTiO_3$ nanopowders containing 5% extra Pb prepared at 460°C for 2 hours, using a) 10°C/min heating rate b) 20°C/min heating rate.

**Figure 6**. The TEM image of the $PbTiO_3$ nanoparticles in our optimized sample, the sample prepared at 460°C for 2 hours.

**Figure 7** The electron diffraction pattern of the prepared nanoparticles in Figure 6 showing single phase crystallization of $PbTiO_3$ tetragonal perovskite phase.

**Tables**

**Table I** Pb content and preparation conditions of different samples which were examined in this study.

| Percentage of crystallization (%) | Firing time (hours) | Heating rate (°C/min) | Firing temperature (°C) | Pb content | Nanopowder samples |
|---|---|---|---|---|---|
| 81.3 | 6 | 20 | 680 | 1.10 | A |
| 72.6 | 6 | 20 | 680 | 1.05 | B |
| 67.5 | 6 | 20 | 680 | 1.00 | C |
| 73.5 | 4 | 20 | 680 | 1.05 | D |
| 78.5 | 2 | 20 | 680 | 1.05 | E |
| 47.5 | 2 | 20 | 460 | 1.05 | F |
| Not calculated | 2 | 10 | 460 | 1.05 | G |

References

1. Beerman HP. Investigation of pyroelectric material characteristics for improved infrared detector performance. Infrared Physics. 1975;15:225-31.
2. Okuyama M, Togami Y, Hamakawa Y. Microwave effect in RF magnetron sputtering of PbTiO$_3$. Applied Surface Science. 1988;33:625-31.
3. Singh V, Suri S, Bamzai K. Mechanical Behaviour and Fracture Mechanics of Praseodymium Modified Lead Titanate Ceramics Prepared by Solid-State Reaction Route. Journal of Ceramics. 2013;2013.
4. TIEN TY, Carlson W. Effect of additives on properties of lead titanate. Journal of the American Ceramic Society. 1962;45:567-71.
5. Ichinose N, Kimura M. Microstructure and piezoelectric properties of modified PbTiO3 ceramics. Japanese journal of applied physics. 1992;31:3033.
6. Szafraniak I, Połomska M, Hilczer B. XRD, TEM and Raman scattering studies of PbTiO3 nanopowders. Crystal Research and Technology. 2006;41:576-9.
7. Wongmaneerung R, Rujiwatra A, Yimnirun R, Ananta S. Fabrication and dielectric properties of lead titanate nanocomposites. Journal of Alloys and Compounds. 2009;475:473-8.
8. Szafraniak-Wiza I, Hilczer B, Talik E, Pietraszko A, Malic B. Ferroelectric perovskite nanopowders obtained by mechanochemical synthesis. Processing and Application of Ceramics. 2010;4:99-106.
9. Chankaew C, Rujiwatra A. Hydrothermal Synthesis of Lead Titanate Fine Powders at Water Boiling Temperature. CHIANG MAI JOURNAL OF SCIENCE. 2010;37:92-8.
10. Lanki M, Nourmohammadi A, Feiz M. A Precise Investigation of Lead Partitioning in Sol-Gel Derived PbTiO3 Nanopowders. Ferroelectrics. 2013;448:123-33.
11. Ishikawa K, Okada N, Takada K, Nomura T, Hagino M. Crystallization and growth process of lead titanate fine particles from alkoxide-prepared powders. Japanese journal of applied physics. 1994;33:3495.
12. Lee C-Y, Tai N-H, Sheu H-S, Chiu H-T, Hsieh S-H. The formation of perovskite PbTiO3 powders by sol–gel process. Materials Chemistry and Physics. 2006;97:468-71.
13. Fang J, Wang J, Gan L-M, Ng S-C. Comparative study on phase development of lead titanate powders. Materials Letters. 2002;52:304-12.

14.     Ishikawa K, Yoshikawa K, Okada N. Size effect on the ferroelectric phase transition in PbTiO 3 ultrafine particles. Physical Review B. 1988;37:5852.

15.     Kakihana M, Okubo T, Arima M, Uchiyama O, Yashima M, Yoshimura M, Nakamura Y. Polymerized complex synthesis of perovskite lead titanate at reduced temperatures: Possible formation of heterometallic (Pb, Ti)-citric acid complex. Chemistry of materials. 1997;9:451-6.

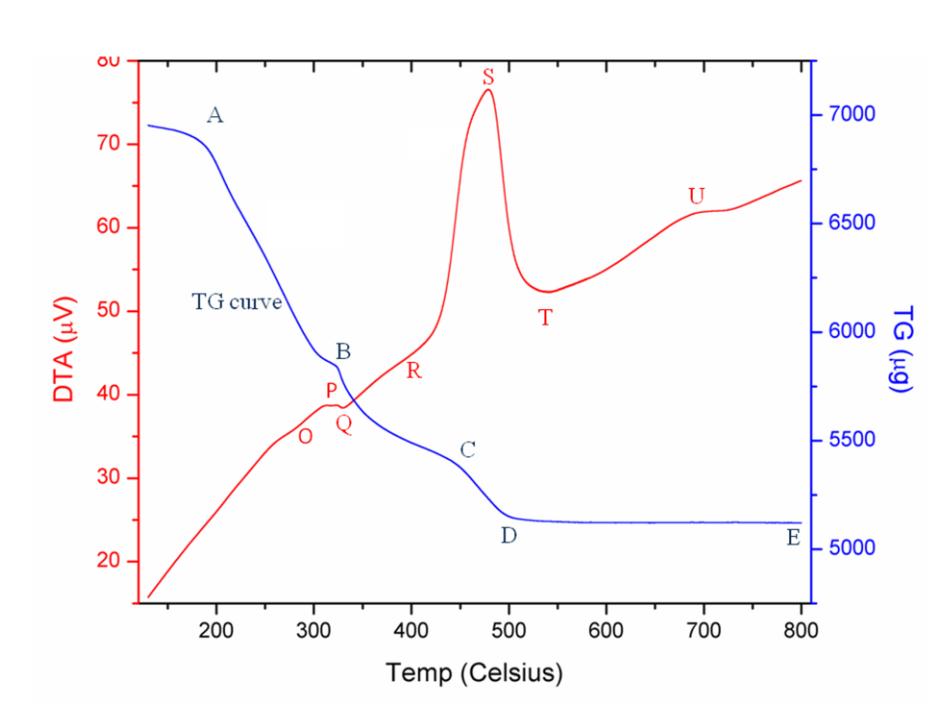

Figure 1

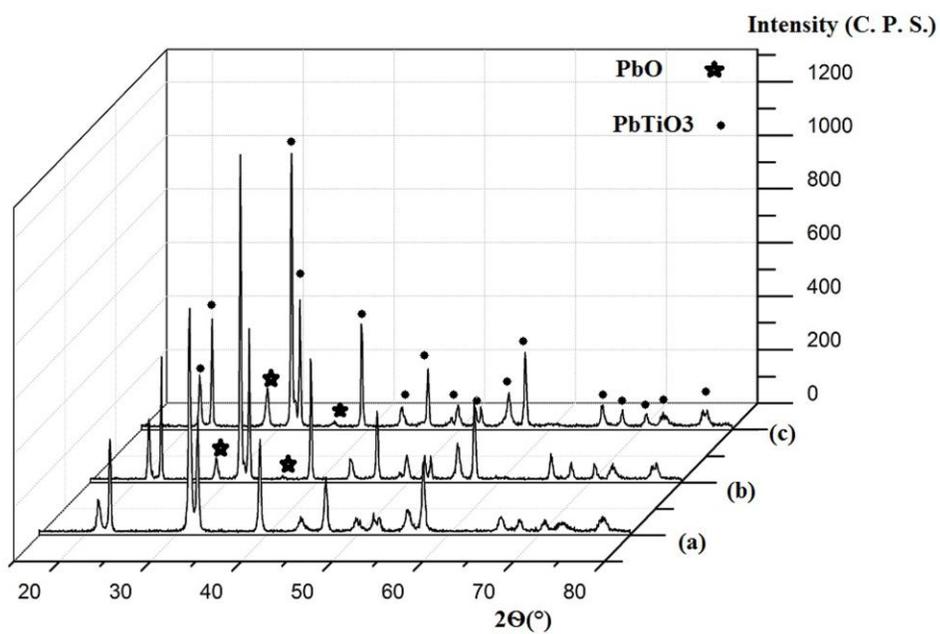

Figure 2

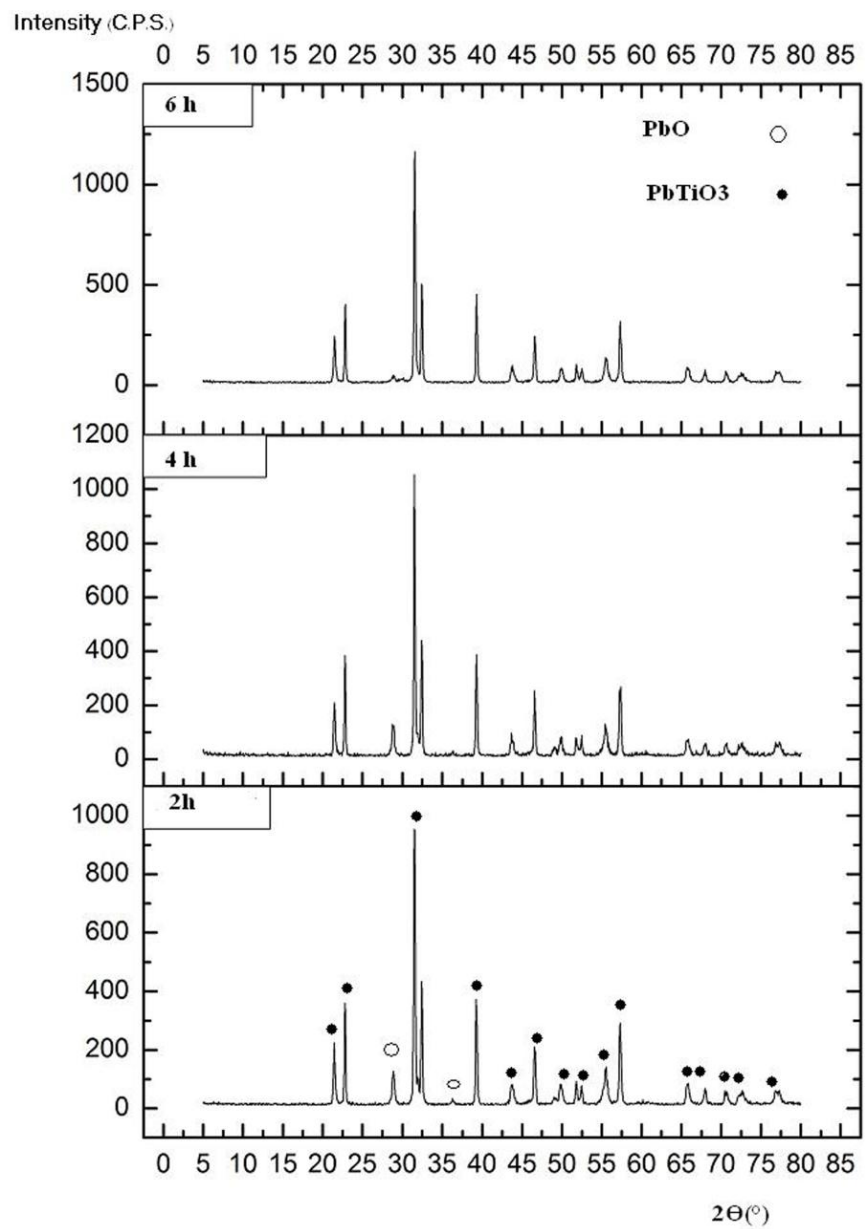

Figure 3

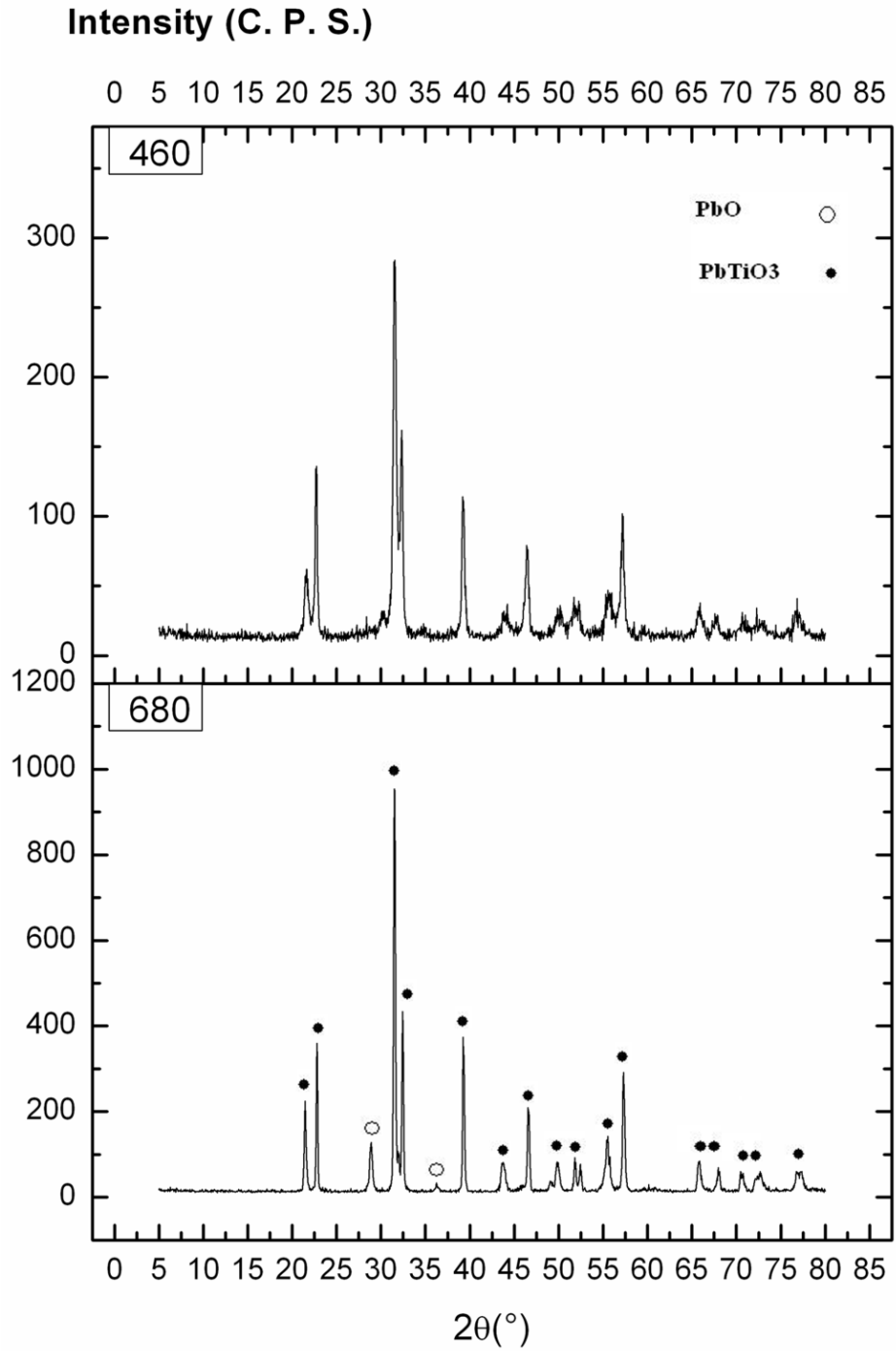

Figure 4

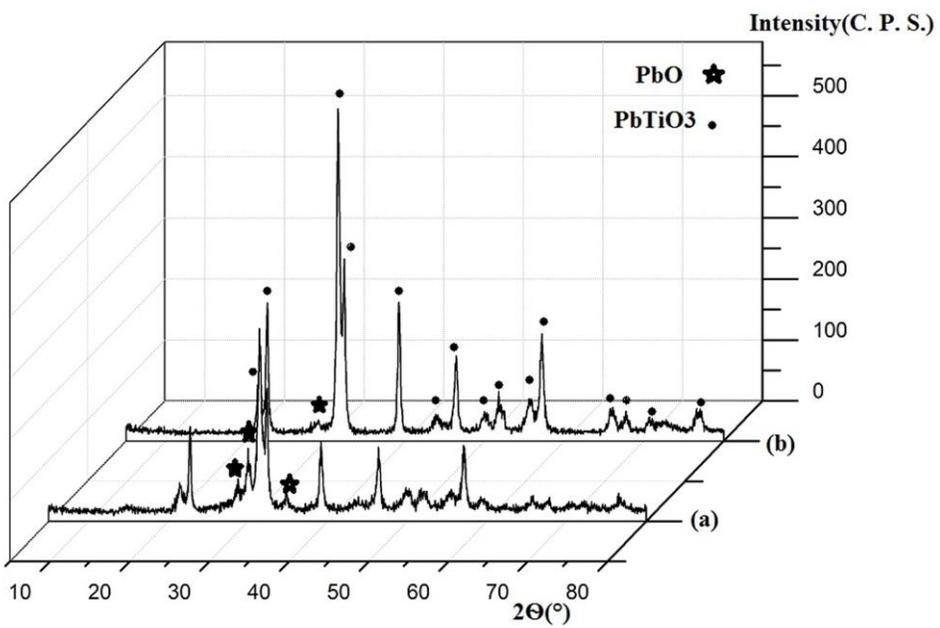

Figure 5

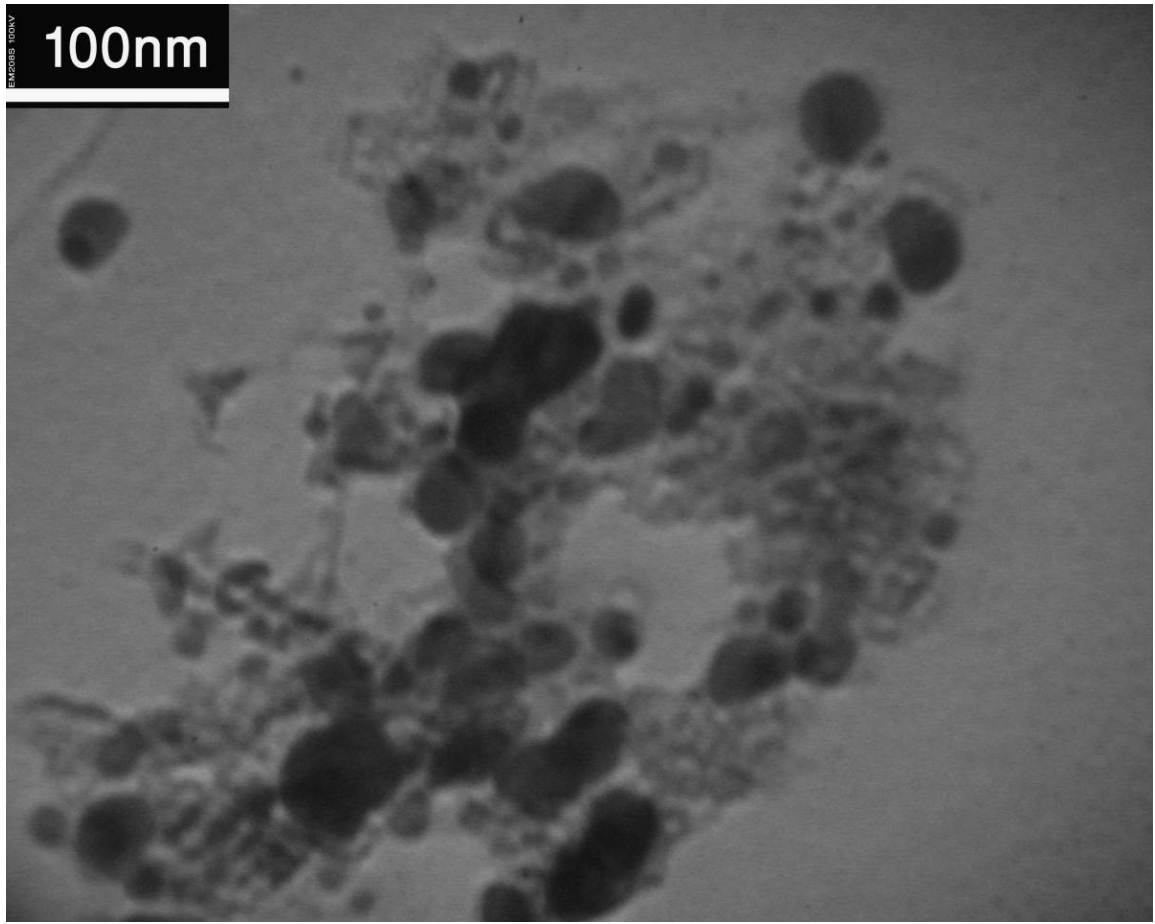

Figure 6

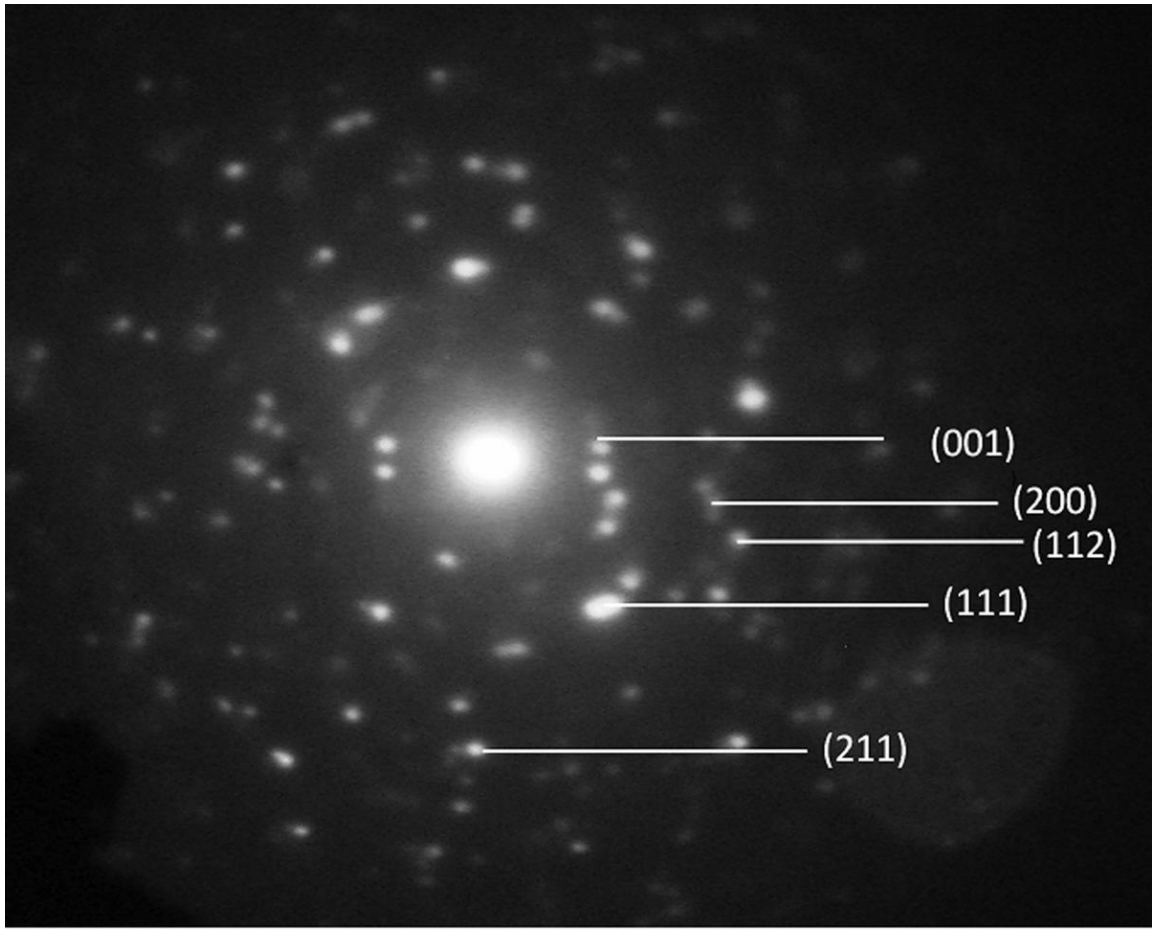

Figure 7